\documentclass[12pt]{article}
\usepackage{epsfig,float,afterpage}

\begin{document}

\begin{large}
\begin{center}
{\Large \bf Search for $CPV$ and New Physics at CLEO\footnote{
Based on invited talks on behalf of CLEO Collaboration 
at international conferences 
Kaon'2001 ($CP$-2001), Pisa,     Italy 
and 
EPS-2001  (HEP-2001),  Budapest, Hungary. 
}}
\end{center}
\end{large}

\vspace{0.1cm}

\begin{center}
        Vladimir Savinov (CLEO Collaboration) \\
        School of Physics and Astronomy, University of Pittsburgh \\
        3941 O'Hara St., Pittsburgh PA USA 15260 \\      
        vps3@pitt.edu, www.phyast.pitt.edu/\~{}savinov
\end{center}

\vspace{0.1cm}

\begin{abstract}
Recent CLEO results on the search for $CP$ violation 
in decays of $B$ and $D$ mesons and $\tau$ lepton 
are reviewed. 
New data on ``wrong-sign'' $D$ decays 
and $B \to K^{(*)}l^+l^-$ FCNC transitions are presented. 
As possible Standard Model contribution to many of 
studied processes is tiny, described efforts 
constitute the search for physics beyond the Standard Model. 
Future CLEO-c efforts on the subject are outlined. 
\end{abstract}

\section{Introduction} 

The ultimate goal of many current and future efforts 
in experimental physics of elementary particles 
is discovering possible new symmetries in Nature. 
These symmetries, commonly referred to as the 
``new physics'' could realize 
via previously unknown interactions 
that we are becoming sensitive to 
with novel detecting devices at collider-based 
experiments as the energies available 
to these machines are growing. 
Besides academic curiosity about 
the processes that might be allowed beyond 
the current energy frontier, 
there are many important questions 
that are outside the scope of 
the Standard Model (SM) of elementary particles.  
Among these are quantifying the imbalance between 
matter and antimatter in visible Universe, 
why there are three generations 
of quarks and leptons in the SM hierarchy, 
what is the connection between 
two fundamental statistics, 
if quarks and leptons are truly point-like 
particles and other open problems. 
Last but not the least, 
I have to mention that 
we have not yet obtained 
the experimental proof of the origin of mass, 
neither we know if the Coulomb's law 
holds at very small distances and 
if there is any connection between gravity 
and possible extra dimensions. 

The results presented in this short review 
were obtained from 
the data collected at the 
Cornell Electron Storage Ring (CESR) 
with the CLEO series of detectors. 
These results are based on statistics that correspond 
to an integrated $e^+e^-$ luminosity of 
up to $9.2 {\rm ~fb^{-1}}$ 
collected at the $\Upsilon(4{\rm S})$ energy 
of 10.58 GeV and up to $4.6 {\rm ~fb^{-1}}$ 
collected approximately 60 MeV below the $\Upsilon(4{\rm S})$ energy. 
These energies are far below the energies available 
at Tevatron, LEP and future LHC experiments, however, 
the tiny effects of new physics at such low energies 
could be amplified by strong interaction 
that might give rise, for example, to direct 
$CP$ asymmetry in particular decays of $B$ and $D$ mesons. 
The measurements described in this review are, most of the time, 
the limits on how rare the searched-for processes are. 
Presenting our experimental results in the form of limits 
on masses, coupling constants and other parameters associated 
with the new physical processes is highly model-dependent 
and is left outside the scope of this review. 
Only more recent CLEO efforts are described, 
while neutrinoless\cite{cleo_tau_neutrinoless}  
and anomalous radiative\cite{cleo_tau_radiative} 
$\tau$ decays, rare $\eta^\prime$ decays\cite{cleo_etaprime_rare}, 
familon\cite{cleo_familon}, 
scalar bottom quark\cite{cleo_squark} 
and other searches 
are {\it not} covered here. 
For some of the processes studied by CLEO  
and reviewed here similar or better limits 
were recently reported 
by BaBar and Belle experiments. 

Our data sample was recorded with two configurations 
of the CLEO detector. 
The first third of the data was recorded with 
the CLEO II detector\cite{CLEOII_description} 
which consisted of three cylindrical drift chambers 
placed in an axial solenoidal magnetic field of 1.5T, 
a CsI(Tl)-crystal electromagnetic calorimeter, 
a time-of-flight (TOF) plastic scintillator system 
and a muon system (proportional counters embedded 
at various depths in the steel absorber). 
Two thirds of the data were taken with 
the CLEO II.V configuration of the detector 
where the innermost drift chamber 
was replaced by a silicon vertex detector\cite{CLEOII.V_description} (SVX) 
and the argon-ethane gas of the main drift chamber 
was changed to a helium-propane mixture. 
This upgrade led to improved resolutions in momentum 
and specific ionization energy loss 
($dE/dx$) measurements. 

The three-tier CLEO trigger system\cite{CLEOII_trigger} complemented by the 
software filter for beam-gas rejection utilized 
the information from the two outer drift chambers, 
the TOF system and electromagnetic calorimeter. 
The response of the detector was modeled with 
a GEANT-based\cite{GEANT} Monte Carlo (MC) simulation program. 
The data and simulated samples were processed 
by the same event reconstruction program. 
Whenever possible the efficiencies were either 
calibrated or corrected for the difference 
between simulated and actual detector responses 
using direct measurements from independent data. 

Most of the measurements reviewed here have been made 
by first measuring signal candidates yields and 
then, when necessary, subtracting background contributions. 
These ``yields'' were not always the numbers of events, 
often these were complicated weighted quantities 
optimized for signal-background separation. 
Simple ``slice and dice'' or counting method 
was employed in simpler analyses, 
more sophisticated techniques, such as 
maximum likelihood (ML) fitting, 
neural nets and optimal observable method 
were applied to more difficult cases. 
The resulting numbers were corrected for 
the overall efficiencies and then converted 
to the measured quantities using luminosities, 
branching fractions and, when necessary, theoretical input. 
Detailed description of these procedures can be found 
in references to CLEO publications. 
The reader can also find relevant theoretical references 
in these publications. 
Our results are based on 
9.6 million $B\bar{B}$ events, 
12.2 million $\tau^+\tau^-$ pairs and 
9 ${\rm fm}^{-1}$ of $e^+e^-$ statistics 
with SVX appropriate for precise $D$ mesons measurements. 

  \section{Bound on $CP$ Asymmetry in $b \to s\gamma$ Decays} 

The theory\cite{kagan_and_neubert} predicts very small ${\cal A}_{CP}$ asymmetry 
in inclusive rate for $b \to s\gamma$ 
where\footnote{Similar definitions of ${\cal A}_{CP}$ are used in all other analyses reviewed here.}  
\begin{equation}
  {\cal A}_{CP} = \frac{{\Gamma}(b \to s \gamma)-{\Gamma}(\bar{b} \to \bar{s} \gamma)}{{\Gamma}(b \to s \gamma)+{\Gamma}(\bar{b} \to \bar{s} \gamma)}.
  \label{eq1} 
\end{equation}
\noindent In SM non-zero ${\cal A}_{CP}$ 
could arise from QCD radiative corrections 
and interference among processes whose amplitudes are 
driven by $c_7$ (QED penguin), $c_2$ (four-fermion) 
and $c_8$ (QCD penguin) Wilson coefficients 
in OEP expansion for effective weak Hamiltonian. 
This asymmetry is proportional to $\alpha_s$ and 
could be enhanced by some new processes that would 
modify effective theory. As with all other 
direct $CP$ asymmetries discussed here, possible 
non-zero ${\cal A}_{CP}$ in $b \to s \gamma$ should vanish 
when summing over {\it all} $B$ (or, separately, over all $\bar{B}$) 
decays is performed. 
This is required to conserve $CPT$ and 
means equal lifetimes for a particle and corresponding antiparticle. 

In our analyses\cite{cleo_acp_bsgamma,cleo_br_bsgamma} 
the signature of $b \to s\gamma$ is high energy photon: 
$2.2~{\rm GeV}~ < E_\gamma < 2.7~{\rm GeV}$. 
To measure partial rates, flavor tagging is necessary. 
We achieved 89\% and 90\% correct $b$-flavor tagging by either 
using a high momentum lepton from the recoiling, 
non-signal $B$ meson candidate, or, when lepton is not identified, 
with the help of pseudoreconstruction technique 
where charged kaon candidate was used for tagging. 
The effects of $B\bar{B}$ mixing were corrected for, 
substantial background from continuum processes 
(including initial state radiation (ISR)) was suppressed using $\pi^0$ and $\eta$ vetoes 
and a combination of kinematic variables provided 
as the input to neural net trained using 
signal MC and background samples. The remaining background was 
subtracted using independent data sample recorded $\approx 60$ MeV 
below $B\bar{B}$ threshold. 

We do not tag strangeness in 
$b \to s\gamma$ (even when using pseudoreconstruction technique) 
because of three reasons: 
there is no kaon requirement in lepton flavor tagging method, 
our PID is far from being perfect in pseudoreconstruction method, 
and, finally, $K\bar{K}$ pairs could be popped up 
from vacuum after $b \to d\gamma$  quark-level transition 
thus faking the $b \to s\gamma$ signal. 
As the result our $b \to s\gamma$ and $b \to d\gamma$ efficiencies 
are almost the same. 
This is acceptable because the theory predicts that 
$b \to d\gamma$ contributes only $\approx 4$\% to the inclusive radiative rate. 
To estimate the $b \to d \gamma$ contribution from data we performed 
searches\cite{cleo_b_dstargamma,cleo_acp_kstargamma} for certain exclusive channels (such as $B \to \rho \gamma$ 
and $B \to D^{*0} \gamma$) 
that would be coming from $B$ decays mediated via QED $b \to d\gamma$ penguin, 
$W$ annihilation and $W$ exchange and did not find these processes 
to have unexpectedly large 
branching fractions. However, our experimental limits on $b \to d \gamma$ 
are still approx. 10 times higher than what we would love to see 
as the experimental proof of $b \to d \gamma$ being small.  
Therefore, we {\it assume} $b \to s\gamma$ after subtracting all known 
contributions. As the result our measured $b \to s\gamma$ asymmetry 
is biased by possible small contribution from $b \to d\gamma$ that 
would give rise to ${\cal A}_{CP}$ asymmetry of opposite (to signal) sign. 

We measure\cite{cleo_acp_bsgamma} 
${\cal A}_{CP} = (-0.079 \pm 0.108 \pm 0.022)(1.0 \pm 0.03)$, 
where the first and second errors  
are statistical and additive systematic (same convention is used for all 
analyses reviewed here), 
and the 3\% error is multiplicative systematic. 
Major contributions to systematics arise from mistag rate, 
continuum and $B\bar{B}$ background subtractions. 
Contribution from particle detection biases (matter-antimatter 
detection efficiency asymmetry) for kaons, pions and leptons 
is very small (less than 1\%) as it is in all other analyses reported here. 
We also established a 90\% confidence level (CL) interval $-0.27 < {\cal A}_{CP} < +0.10$ 
ruling out some extreme (though unspecified) non-SM predictions. 

  \section{Exclusive QED Penguins and $CP$ Asymmetry} 

Prompted by theoretical expectation of non-zero ${\cal A}_{CP}$ 
in inclusive $b \to s\gamma$ CLEO also measured this asymmetry  
in exclusive channels $B \to K^*(892) \gamma$. 
In this case flavor self-tagging was achieved using 
charged kaon or pion 
and it was assumed that ${\cal A}_{CP}$ is the same for neutral and 
charged $B$ mesons depending only on $b$ quark flavor. 
We measured\cite{cleo_acp_kstargamma} 
${\cal A}_{CP} = +0.01 \pm 0.06$, where only statistical 
error is shown (in this review we usually do not show systematics 
for statistics-limited measurements). 
To obtain this number we corrected for misidentification 
rate of $\approx 3.5$\%. 

  \section{$CP$ Asymmetry in Dileptons from $B\bar{B}$ Decays} 

Same-sign dileptons appear naturally in $\Upsilon(4S) \to B^0 \bar{B^0}$ 
decays because of $B\bar{B}$ mixing. Similarly to neutral kaon decays, 
$B$ mass and flavor eigenstates could be different with the former 
described in terms of the latter 
as $[(1+\epsilon_B)B^0+(1-\epsilon_B)\bar{B}^0]/\sqrt{2(1+|\epsilon_B|^2)}$. 
This could give rise to $CP$ violation ($CPV$) that would 
manifest itself in a non-zero value for 
\begin{equation}
  a_{ll} = \frac{N(l^+l^+)-N(l^-l^-)}{N(l^+l^+)+N(l^-l^-)} \approx \frac{4 {\cal R }e(\epsilon_B)}{1+|\epsilon_B|^2}, 
  \label{eq2} 
\end{equation}
\noindent where index $l$ stands for a lepton from semileptonic $B$ decay. 

Experimental challenges in this analysis include measuring 
dilution factor and the charge asymmetry in the fake probability 
from the data, 
continuum suppression and subtraction. Only high-momentum leptons are 
used in the measurement: $1.6 ~{\rm GeV/c}~< p_l < 2.4 ~{\rm GeV/c}$. 
In this analysis\cite{cleo_b_dileptons} we measured 
$a_{ll} = (+0.013 \pm 0.050 \pm 0.005)(1.00 \pm 0.10)$. 
By combining this result with our previous 
(statistically independent) measurement\cite{cleo_b_mixing} 
(where $B\bar{B}$ mixing was the main subject) we arrive at weighted average 
${\cal R}e(\epsilon_B)/(1+|\epsilon|^2) = +0.0035 \pm 0.0103 \pm 0.0015$. 
This is significant improvement in comparison to CDF 
result\cite{cdf_b_dileptons}: $+0.025 \pm 0.062 \pm 0.032$. 
Our sensitivity is similar 
to that of OPAL's result\cite{opal_b_dileptons}: 
$+0.002 \pm 0.007 \pm  0.003$. 
In contrast to these two measurements our analysis 
is not affected by {\it possible} $B_s$ contamination 
as these mesons can not be produced at energies 
available to us at $\Upsilon(4S)$. 

  \section{Search for $CP$ Violation in $B \to \psi^{(\prime)} K$ Decays} 

Possible SM contribution to direct $CP$ asymmetry in 
$B^\pm \to \psi^{(\prime)} K^\pm$ decays is practically zero. 
However, some SM extensions would be able to explain 
non-zero ${\cal A}_{CP}$ (if observed). These are 2HDM 
extensions\cite{soni} where $t$ quark plays a special role 
and masses of up- and down-type fermions are generated 
by different Higgs doublets. 
Presence of dileptons from $\psi^{(\prime)}$ decays 
allows a background-free analysis and we measured\cite{cleo_acp_jpsi} 
${\cal A}_{CP}(B^\pm \to \psi K^\pm) = +0.018 \pm 0.043 \pm 0.004$ 
and 
${\cal A}_{CP}(B^\pm \to \psi^\prime K^\pm) = +0.02 \pm 0.091 \pm 0.01$. 
These two channels are not combined together because 
rescattering that would give rise to FSI and $CP$-conserving 
strong phase necessary for observing the effects of ($CP$ non-conserving) 
new physics is channel-dependent. 

  \section{$CP$ Asymmetries in Charmless Hadronic $B$ Decays} 

We also measured\cite{cleo_acp_charmless} ${\cal A}_{CP}$ for five 
two-body charmless hadronic $B$ decays. The theory\cite{ali_cp} 
predicts ${\cal A}_{CP} < 0.1$ for these decays assuming factorization 
and no soft FSI. 
The non-zero SM ${\cal A}_{CP}$ arises from the interference 
between tree $b \to u$ and penguin $b \to s$ transitions. 
There are indications that their amplitudes in studied $B$ decays 
are comparable, therefore (assuming large strong phase) 
${\cal A}_{CP}$ might be observable with existing data. 
Once again, 
if there is new physics contribution in the quark-level 
transitions for decays of heavy flavors and FSI 
(or rescattering) is large, ${\cal A}_{CP}$ 
could be much larger. 
Using self-tagging we found 
${\cal A}_{CP}(B \to K^\pm \pi^\mp) = -0.04 \pm 0.16$, 
${\cal A}_{CP}(B \to K^\pm \pi^0)   = -0.29 \pm 0.23$, 
${\cal A}_{CP}(B \to K^0_s \pi^\pm) = +0.18 \pm 0.24$, 
${\cal A}_{CP}(B \to K^\pm \eta^\prime) = +0.03 \pm 0.12$ 
and 
${\cal A}_{CP}(B \to \omega \pi^\pm) = -0.34 \pm 0.25$, 
where only statistical errors are shown. 
The interpretation of these results, 
consistent with the SM predictions, 
would be subject to FSI contribution 
uncertainties even if some ${\cal A}_{CP}$ 
were established to be non-zero. 
As in many other analyses where SM $CPV$ {\it is} allowed, 
it is strong interaction that would make (possible) new physics effects 
to realize at low energies, while quantifying its role 
in extracting parameters associated with these effects 
could be very difficult. 

  \section{Search for $CP$ Violation in $\tau$ Decays} 

In contrast to charmless $B$ decays, 
no SM $CP$ violation is possible in $\tau$ sector and 
our previous search\cite{cleo_tau_old} did not find any 
in our $\tau^\pm \to K^0_s \pi^\pm \nu$ data. 
Since than we realized that a more powerful search 
is possible using the optimal observable method\cite{atwood_optimal}. 
This method is based on presenting the data in the form 
most sensitive to a particular SM extension. 
We did not find $CP$ violation in our new $\tau^\pm \to \pi^\pm \pi^0 \nu$ 
analysis\cite{cleo_tau_new} and, analyzing our data 
using a particular 3HDM SM extension\cite{grossman_lambda} 
established a 90\% CL 
interval on the parameter $\Lambda$ that describes convoluted 
complex combination of this model's coupling constants: 
$-0.046 < {\cal I}m (\Lambda) < 0.022$. 
In this model $\tau$ can also decay via charged Higgs boson. 
We also reanalyzed $\tau^\pm \to K^0_s \pi^\pm \nu$ 
data in the same spirit and measured\cite{cleo_tau_newest} 
$-0.155 < {\cal I}m (\Lambda) < 0.047$ 90\% CL interval 
(preliminary result). 
Our new results are obtained using conservative 
form-factor model for hadronic part of studied $\tau$ decays. 
Namely, we experimented with three models for 
form factors and chose the one that generated smallest strong phase 
(thus reducing our potential sensitivity to hidden new physics). 

  \section{$CP$ Violation in $D$ Decays to Pseudoscalar Pairs} 

We measured ${\cal A}_{CP}$ for five $D^0$ decay channels: 
${\cal A}_{CP}(D \to K^+ K^-) = +0.0005 \pm 0.0218 \pm 0.0084$ 
and 
${\cal A}_{CP}(D \to \pi^+ \pi^-) = +0.020 \pm 0.032 \pm 0.008$ 
(preliminary results\cite{cleo_d_proper_time}), 
${\cal A}_{CP}(D \to K^0_s \pi^0) = +0.001 \pm 0.013$, 
${\cal A}_{CP}(D \to \pi^0 \pi^0) = +0.001 \pm 0.048$ 
and 
${\cal A}_{CP}(D \to K^0_s K^0_s) = -0.23 \pm 0.19$, 
where only statistical errors are shown 
for the last three channels\cite{cleo_acp_d0}. 
In these analyses $d$ flavor 
was tagged by the charge of pion from 
two-body decay $D^{*+} \to \pi^+ D^0$ (and charge-conjugate). 
Assuming that $D^*$ decay is $CP$-conserving, 
to measure ${\cal A}_{CP}$ for $D$ mesons we actually measure 
the numbers of $D^{*+}$ ($D^{*-}$) decays 
followed by the $D^0$ ($\bar{D^0}$) decay. 
Possible non-zero ${\cal A}_{CP}$ would be manifestation of 
direct $CP$ violation which is usually predicted 
to be small in SM, of the order of 0.1\% 
in the charmed mesons sector\cite{bucella}. 
However there are indications that 
large FSI effects are present in $D$ decays and this would 
make them an excellent place to search for $CP$ violation 
with more data expected at CLEO-c\cite{cleoc}. 

  \section{New Fits to $D^0$ Proper Time and $y$ Measurement} 

The first two ${\cal A}_{CP}$ measurements for charmed mesons 
described above are actually a by-product of the new 
analysis of $D^0$ proper time measured 
with CLEO II.V SVX by reconstructing the 
displaced vertices of the $D^0$ mesons decaying in flight. 
Proper time measurements provide us 
with information on $D\bar{D}$ mixing 
that can proceed via on-shell and off-shell 
intermediate states. In a widely accepted 
framework the amplitude for former (latter) 
contribution is $-iy$ ($x$) 
in units of $\Gamma_{D^0}/2$. 
The signatures of new physics include 
$|x| >> |y|$ and possible large $|{\cal I}m(x)/x|$ causing 
$CP$ violating interference between $x$ and $y$ or between 
$x$ and direct decay amplitudes to the same final state. 
SM $x$ contribution to $D\bar{D}$ mixing is strongly Cabibbo-suppressed: 
$|x| \approx \tan^2{\theta_C} \approx 5$\% and this 
is reduced further down to $|x| \approx 10^{-6}$--$10^{-2}$ 
by GIM cancelation where large uncertainty is due to 
strong interaction effects. 
In the limit of no $CP$ violation in $D^0$ decays, 
one can measure $y$ by comparing distributions of decay distances 
({\it i.e.} by measuring proper times) 
for samples of $D^0$ mesons decaying to $CP$ eigenstates 
and to the states of mixed $CP$. We measured\cite{cleo_d_proper_time} 
proper times for $D^0$ decays to $K^+K^-$, $\pi^+\pi^-$ and 
$K^+\pi^-$ final states and obtained preliminary result 
$y = -0.011 \pm 0.025 \pm 0.014$ which is consistent with zero. 
This is in agreement  
with our previous measurement\cite{cleo_d_proper_time_old} 
and recent FOCUS results\cite{focus_d_proper_time}. 

  \section{First Measurement of $D^0 \to K^+\pi^-\pi^0$ Rate} 

Charmed meson initially tagged as $D^0$ 
can be observed in  a ``wrong-sign'' (WS) final state, 
such as $K^+\pi^-\pi^0$ as the result of 
direct doubly Cabibbo-suppressed decay (DCSD) 
or via a ``right-sign'' (RS) $\bar{D^0}$ decay 
after mixing. In this analysis\cite{cleo_d_ws_kpp} we measure 
time-integrated ratio 
\begin{equation}
R(K\pi\pi) = \frac{\Gamma(D^0 \to K^+\pi^-\pi^0)}{\Gamma(\bar{D^0} \to K^+\pi^-\pi^0)} = R_D(K\pi\pi) + \sqrt{R_D(K\pi\pi)} y^\prime + \frac{x^{\prime 2}+y^{\prime 2}}{2}, 
  \label{eq3} 
\end{equation}
\noindent where $R_D(K\pi\pi)$ is the relative rate of DCSD for this channel, 
$x^\prime = x \cos{\delta} + y \sin{\delta}$, 
$y^\prime = y \cos{\delta} - x \sin{\delta}$ and 
$\delta$ is a possible strong phase between 
direct and mixing-induced WS amplitudes.  

We measure $R(K\pi\pi)$ from the 
two-dimensional ML fit in $m(K\pi\pi)$ {\it vs.} $Q = M_{D^*} - M_{K\pi\pi} - m_\pi$ 
and use the results to establish the WS decay. 
From our binned ML fit we measure $R(K\pi\pi) = 0.0043^{+0.0011}_{-0.0010} \pm 0.0007$. 
With more statistics we could have made a time-dependent fit to extract $x^\prime$, 
$y^\prime$ and $R_D(K\pi\pi)$ simultaneously (although with correlations), 
however, with our data (observing $38 \pm 9$ WS $K\pi\pi$ events) we 
can only measure $R_D(K\pi\pi) = (1.7 \pm 0.4 \pm 0.3) \times \tan^4{\theta_C}$ 
assuming no mixing. If, instead, we assumed $\delta(K\pi\pi) = \delta(K\pi)$ (unfounded), 
$y^\prime=0$ and $|x^\prime|=0.028$ 
(using 95\% CL upper limit from 
our previous measurement\cite{cleo_d_proper_time_old}), 
these changes would have had 
very small effect on our $R_D(K\pi\pi)$ result. 

  \section{New Results on FCNC Decays $B \to K^{(*)}l^+l^-$} 

Processes mediated by (effective, in SM) 
Flavor Changing Neutral Currents (FCNC) 
are among the best candidates to 
search for new physics in $B$ decays. 
Field-theoretical description of FCNC decays 
contains loop and box diagrams and this 
would make it easy to explain deviations 
from SM (if observed) by introducing 
much heavier non-SM particles 
that would affect the total rate 
and 
modify differential distributions 
for these decays. 

Background suppression is one of the challenges present 
in this analysis. In addition to continuum (includes ISR) 
there are dileptons from $\psi^{(\prime)}$ and $B\bar{B}$ decays. 
Almost all background-suppression variables and 
methods developed on CLEO are combined to achieve the 
best possible sensitivity in our blind analysis that uses 
12 $e^+e^-$ and $\mu^+\mu^-$ exclusive channels. 
To make $\psi^{(\prime)}$ suppression efficient 
we correct their four momenta to include 
internal and external bremsstrahlung photons (when recovered) 
and apply wide vetoes on invariant masses. 
For example, in $e^+e^-$ channels we veto candidates 
with (corrected, when applicable) dielectron invariant masses 
in the range between 2.80 and 3.23 GeV. 
To suppress $B\bar{B}$ dileptons 
we employ missing mass technique developed on CLEO for 
neutrino ``detection''. 
Careful studies of $e$ and $\mu$ reconstruction efficiencies 
using data have been carried over the past two years and 
this is the reason why our new FCNC 
results became available only recently. 
To suppress virtual $K^*\gamma$ contribution 
we select events with dilepton invariant mass $m_{ll} > 0.5 {\rm ~GeV/c^2}$ 
in the $K^*$ analyses. Continuum 
background is suppressed using Fisher discriminant 
that combines all available kinematic information 
about each $B$ candidate and the event as a whole. 
In each channel selection criteria were optimized 
for the discovery and best upper limit 
with the average used in the actual analysis. 
This gave us overall (weighted) efficiency of $54.8$\%. 

Summing over all channels 
we counted the number of observed events in the signal 
box of beam-constrained $B$ mass {\it vs.} $\Delta E$, 
the difference between the beam energy and that of a candidate 
and found 7 events while our data-based estimates of 
background contribution is $5.8 \pm 0.8$. 
As the result we achieved excellent sensitivity 
and measured\cite{cleo_b_kll_new} 90\% CL (Bayesian) upper limits 
for $B$ branching fractions 

\begin{center}
${\cal B}(B \to K l^+l^-) < 1.7 \times 10^{-6}$, \\
${\cal B}(B \to K^*(892) l^+l^-)_{m_{ll}>0.5 {\rm ~GeV}} < 3.2 \times 10^{-6}$ and \\
$[0.65 {\cal B}(B \to K l^+l^-) + 0.35 {\cal B}(B \to K^*(892) l^+l^-)_{m_{ll}>0.5 {\rm ~GeV}}] < 1.5 \times 10^{-6}$, 
\end{center}

\noindent where the latter efficiency-weighted limit is just 50\% 
above the SM prediction\cite{ali_kll} with almost no model 
dependence in the efficiency. We estimate that 
to become actually sensitive to new physics contribution in these decays 
data samples of the order of $500 {\rm ~fb^{-1}}$ might be necessary. 

  \section{Future $\tau$-charm Factory at Cornell} 

At future CLEO-c experiment\cite{cleoc} we plan to use 
the existing CLEO III apparatus with silicon vertex detector 
replaced by a compact low-mass inner drift chamber.  
This detector will collect data at the $J/\psi$ and 
$e^+e^- \to D\bar{D}$ threshold energies with initial CESR-c 
instantaneous luminosity around $3$--$5 \times 10^{32} {\rm ~cm^{-2}~s^{-1}}$. 
CLEO-c will operate with reduced magnetic field (down to 1.0 Tesla
from 1.5 Tesla at present, this only requires turning a knob)
in the main part of the detector. 
At CLEO-c we plan to collect samples of 1.5 million ${D_s\bar{D_s}}$ pairs, 
30 million $D\bar{D}$ pairs and 1 billion $J/\psi$ decays. 
These exceed the MARK-III statistics by the factors of 
480, 310 and 170, respectively. 

An important element of our CLEO-c program is to measure 
the $D$ mesons coupling constants that are important for 
testing Lattice QCD (LQCD) predictions. If these predictions are confirmed 
by the measurements, there would be a degree of confidence in LQCD 
calculations for $f_B$. Then the theory could be used to 
reduce hadronic uncertainties in future measurements of CKM 
matrix elements at $B$-factories and Fermilab. 
Reducing hadronic uncertainties in CKM measurements 
should help to ``over-constrain'' 
the unitarity triangles thus possibly leading to hints of 
new physics contribution to the decays of heavy mesons. 
The measurements of $f_{D_{(s)}}$ coupling constants will be done 
at CLEO-c by precisely measuring the rates for the $D \to l \bar{\nu}$ 
and $D_s \to l \bar{\nu}$ decays. 
The main advantage of CLEO-c will be the ability to ``tag'' 
the signal $D$ meson decaying leptonically by fully reconstructing 
the recoiling $D$ and by detecting a lepton from signal $D$ decay. 
Applying kinematic constraints will help to bring the 
neutrino ``detection'' technique to a qualitatively better level 
than is possible on CLEO at present in $B$ decays. 
According to our estimates we should be able to achieve 
the 2\% precision in the measurements of $f_D$ and $f_{D_s}$. 
At present these are known with 35\% and 100\% errors, respectively. 
We also plan to measure many other $D$ branching fractions 
with much better precision than these are known at present. 
Knowledge of these branching fractions, such as ${\cal B}(D \to K\pi)$ 
is important for normalizing measurements at higher energies. 

In our future $D\bar{D}$ mixing and $CP$ asymmetries studies 
we should benefit from the fact that $D$ and $\bar{D}$ 
at CLEO-c will be produced in a coherent quantum state. 
This is crucial for direct $CP$ searches we plan to undertake. 
The key point is that when a $D\bar{D}$ pair is produced in a 
$C$-odd initial state from $e^+e^-$ annihilation, 
a final state with two $CP$ eigenstates of the same $CP$ eigenvalue 
would be an unambiguous manifestation of direct $CP$ violation. 
As the result of our CLEO-c program we expect 
to achieve the precision of better than 1\% 
in the measurements of $D\bar{D}$ mixing and 
${\cal A}_{CP}$ asymmetry parameter for $D$ mesons. 

The decays of $J/\psi$ will be studied at CLEO-c 
with an unprecedented precision and in much detail. 
Our main emphasis here would be on studying radiative decays 
of $J/\psi$ to light mesons as these provide a glue-rich 
experimental laboratory where new states of 
hadronic matter, generally referred to as glueballs and exotics 
could be finally unambiguously discovered and studied. 
With one billion $J/\psi$ decays we will be able 
to perform partial wave analyses and global fits for 
many mysterious decays of $J/\psi$ (such as $J/\psi \to \gamma f_j(2220)$) 
presumably observed previously, however, with poor statistics. 

The measurements of $R$, the total hadronic cross section 
in $e^+e^-$ annihilation in a wide range of $e^+e^-$ invariant masses 
($\sqrt{S}$) proposed at CLEO-c is an integral part of global efforts 
in heavy flavor physics. These measurements 
will help to reduce uncertainties associated with extracting 
new physics effects from existing and future data at higher energies 
experiments. 
By measuring ${\cal R}$ in a wide range of $\sqrt{S}$ 
we should be able to further constrain 
the mass of the (minimal) SM Higgs. 
In this case, the analysis of ${\cal R}$ 
would result in the measurement of running $\alpha$(electromagnetic) 
that should allow to put more restrictive limits on the mass of SM Higgs 
if it is not yet discovered by the time CLEO-c data is available. 
Finally, precise scan of ${\cal R}$ might give information 
about vector hybrids, {\it i.e.} hadrons consisting of $q\bar{q} g$ whose existence 
is predicted by LQCD calculations. The latter project's feasibility 
would depend on the magnitudes of $e^+e^-$ partial widths of such 
hypothetic resonances. 
We expect to achieve the 2\% precision per point 
in the measurements of $R$ at CLEO-c 
in the invariant mass region between 3 and 5 GeV. 
CESR-c will have the ability to run with a single beam energy 
of up to 5.6 GeV thus opening the prospects to do the measurements 
at even higher $\sqrt{S}$. As relevant for $R$ measurements 
we are especially interested in $\sqrt{S}$ 
region above $J/\psi$ and below $\Upsilon$ resonances. 
CESR-c and CLEO-c is a three-year program that is expected to start 
in the end of 2002. 

  \section{Conclusions and Acknowledgments} 

CLEO has searched for a variety of $CP$ asymmetries, 
``wrong-sign'' $D^0$ decays and $B$ decays mediated by 
FCNC transitions with the purpose to probe the 
non-SM physics. No evidence of such contribution was 
found. More precise measurements are starting coming from 
other $B$ physics experiments, better limits on or the discovery 
of $D\bar{D}$ mixing and $CP$ violation in $D$ decays should be 
expected at these facilities and at future $\tau$-charm factory 
at Cornell (CLEO-c and CESR-c). 

I would like to thank the CESR staff for providing us with excellent luminosity 
and my CLEO colleagues who gave me a chance to represent our experiment. 



\end{document}